\documentclass[sigconf, nonacm]{acmart}
\usepackage{amsthm}
\usepackage{amsmath}
\usepackage{commath}
\usepackage{svg}
\usepackage{caption}
\usepackage{subcaption}

\theoremstyle{definition}
\newtheorem{observation}{Observation}
\newcommand\vldbdoi{XX.XX/XXX.XX}
\newcommand\vldbpages{XXX-XXX}
\newcommand\vldbvolume{14}
\newcommand\vldbissue{1}
\newcommand\vldbyear{2020}
\newcommand\vldbauthors{\authors}
\newcommand\vldbtitle{\shorttitle} 
\newcommand\vldbavailabilityurl{https://github.com/eliotwrobson/CyNetDiff/blob/main/examples/visualizations.ipynb}
\newcommand\vldbpagestyle{plain}
\newcommand{\cynetdiff}{\textsc{CyNetDiff}}
\newcommand{\ndlib}{\textsc{NDlib}}

\usepackage{hyperref}
\usepackage{amsmath}
\usepackage{amsthm}

\usepackage{algorithm}
\usepackage{algpseudocode}
\usepackage{multirow}
\usepackage{multicol}
\usepackage{mathtools}

\def\BibTeX{{\rm B\kern-.05em{\sc i\kern-.025em b}\kern-.08em
    T\kern-.1667em\lower.7ex\hbox{E}\kern-.125emX}}

\usepackage[utf8]{inputenc}
\usepackage{pgfplots}
\usepgfplotslibrary{groupplots,dateplot}
\usetikzlibrary{patterns,shapes.arrows}
\pgfplotsset{compat=newest}
\pgfplotsset{scaled y ticks=false}
\usepackage{caption}
\usepackage{subcaption}
\begin{document}
\title{\cynetdiff{}: A Python Library for Accelerated Implementation of Network Diffusion Models}

%%
%% The "author" command and its associated commands are used to define the authors and their affiliations.
\author{Eliot W. Robson}
\affiliation{%
  \institution{University of Illinois Urbana-Champaign}
  %\streetaddress{P.O. Box 1212}
  \city{Urbana}
  \state{Illinois}
  \postcode{61801}
}
\email{erobson2@illinois.edu}

\author{Dhemath Reddy}
\affiliation{%
  \institution{University of Illinois Urbana-Champaign}
  %\streetaddress{P.O. Box 1212}
  \city{Urbana}
  \state{Illinois}
  \postcode{61801}
}
\email{dhemath2@illinois.edu}

\author{Abhishek K. Umrawal}
\affiliation{%
  \institution{University of Illinois Urbana-Champaign}
  %\streetaddress{P.O. Box 1212}
  \city{Urbana}
  \state{Illinois}
  \postcode{61801}
}
\email{aumrawal@illinois.edu}

% \orcid{0000-0002-1825-0097}
% \affiliation{%
%   \country{Iceland}

% \orcid{0000-0001-5109-3700}
% \affiliation{%
% \email{vb@rocquencourt.com}

% \author{J\"org von \"Arbach}
% \affiliation{%
%   \institution{University of T\"ubingen}
%   \city{T\"ubingen}
%   \country{Germany}
% }
% \email{jaerbach@uni-tuebingen.edu}
% \email{myprivate@email.com}
% \email{second@affiliation.mail}

% \author{Wang Xiu Ying}
% \author{Zhe Zuo}
% \affiliation{%
%   \institution{East China Normal University}
%   \city{Shanghai}
%   \country{China}
% }
% \email{firstname.lastname@ecnu.edu.cn}

% \author{Donald Fauntleroy Duck}
% \affiliation{%
%   \institution{Scientific Writing Academy}
%   \city{Duckburg}
%   \country{Calisota}
% }
% \affiliation{%
%   \institution{Donald's Second Affiliation}
%   \city{City}
%   \country{country}
% }
% \email{donald@swa.edu}

\begin{abstract}
In recent years, there has been increasing interest in network diffusion models and related problems. The most popular of these are the independent cascade and linear threshold models. Much of the recent experimental work done on these models requires a large number of simulations conducted on large graphs, a computationally expensive task suited for low-level languages. However, many researchers prefer the use of higher-level languages (such as Python) for their flexibility and shorter development times. Moreover, in many research tasks, these simulations are the most computationally intensive task, so it would be desirable to have a library for these with an interface to a high-level language with the performance of a low-level language. To fill this niche, we introduce \cynetdiff{}, a Python library with components written in Cython to provide improved performance for these computationally intensive diffusion tasks.
\end{abstract}

\maketitle

\if 0
%% do not modify the following VLDB block %%
%% VLDB block start %%%
\pagestyle{\vldbpagestyle}
\begingroup\small\noindent\raggedright\textbf{PVLDB Reference Format:}\\
\vldbauthors. \vldbtitle. PVLDB, \vldbvolume(\vldbissue): \vldbpages, \vldbyear.\\
\href{https://doi.org/\vldbdoi}{doi:\vldbdoi}
\endgroup
\begingroup
\renewcommand\thefootnote{}\footnote{\noindent
This work is licensed under the Creative Commons BY-NC-ND 4.0 International License. Visit \url{https://creativecommons.org/licenses/by-nc-nd/4.0/} to view a copy of this license. For any use beyond those covered by this license, obtain permission by emailing \href{mailto:info@vldb.org}{info@vldb.org}. Copyright is held by the owner/author(s). Publication rights licensed to the VLDB Endowment. \\
\raggedright Proceedings of the VLDB Endowment, Vol. \vldbvolume, No. \vldbissue\ %
ISSN 2150-8097. \\
\href{https://doi.org/\vldbdoi}{doi:\vldbdoi} \\
}\addtocounter{footnote}{-1}\endgroup

\ifdefempty{\vldbavailabilityurl}{}{
\vspace{.3cm}
\begingroup\small\noindent\raggedright\textbf{PVLDB Artifact Availability:}\\
The source code, data, and/or other artifacts have been made available at \url{\vldbavailabilityurl}.
\endgroup
}
%% VLDB block end %%%
\fi

\textbf{Artifact Availability:} \cynetdiff{}---the Python library introduced in this paper is available at https://pypi.org/project/cynetdiff/. The source code, data, and/or other artifacts for this paper are available at https://github.com/eliotwrobson/CyNetDiff/blob/main/examples. Refer to the Jypyter notebook titled visualizations.ipynb for a demonstration of the performance of \cynetdiff{}.

\section{Introduction}
\noindent \textit{Motivation.} Network diffusion is central to studying information propagation \cite{umrawal2023community, umrawal2023fractional} and epidemic spreading \cite{burkholz2021cascade} over social networks. There are several discrete-time stochastic models of diffusion over social networks. In this work, we focus on the \textit{independent cascade} (IC) \cite{goldenberg2001talk, goldenberg2001using} and \textit{linear threshold} (LT) \cite{granovetter1978threshold, schelling2006micromotives} models of diffusion (discussed in Section \ref{sec:preliminaries}). In particular, many research tasks related to these models involve simulating their execution over large networks. This can become computationally expensive as both the size of the graphs and the number of simulations grow. Of particular note is the task of influence maximization (IM), introduced by Domingos and Richardson \cite{domingos2001mining}. This involves selecting users on social networks to sponsor to maximize influence under some network diffusion models. This has been widely studied in different settings, namely the discrete \cite{kempe2003maximizing,umrawal2023leveraging,umrawal2023community}, continuous \cite{chen2020scalable,umrawal2023fractional}, and online \cite{agarwal2022stochastic,nie2022explore}. Many algorithms for this task involve computing the influence of a given set under a network diffusion model that requires a large number of simulations.

\noindent \paragraph{Related Work.}
Other Python libraries have been written for this task, most notably \ndlib{} \cite{rmrspg-ndlib-18}. This is a library that can simulate various network diffusion models (including the IC and LT models described here), written in pure Python, on top of the NetworkX graph library \cite{hss-networkx-08}. \ndlib{} suffers from shortcomings inherent to many pure Python libraries, including large memory overhead, and slower execution of iterative algorithms than a compiled language. In addition, \ndlib{} simulates these models by looping through every node in each time step of the model, meaning that this computation is inefficient when only a few nodes are active.

\noindent \paragraph{Contribution.}
In this work, we introduce the \cynetdiff{} library for simulating these diffusion tasks.
We describe the technologies, data structures, and algorithms used in the implementation. We demonstrate this greater performance with a detailed set of benchmarks. We apply this library to influence maximization by implementing the CELF algorithm \cite{leskovec2007cost} and reporting the performance.

% TODO add section references here
% \begin{enumerate}
%     \item We describe the technologies, data structures, and algorithms used in the implementation of \cynetdiff{}.

%     \item We demonstrate this greater performance with a detailed set of benchmarks.

%     \item We provide an application of this library to influence maximization by implementing the CELF algorithm and reporting the performance.
% \end{enumerate}

\noindent \paragraph{Organization.}
The rest of this paper is organized as follows. In Section \ref{sec:preliminaries}, we discuss the background for the use cases served by \cynetdiff{}. In Section \ref{sec:cynetdiff}, we discuss the implementation details of the package and how it is optimized for the previously discussed use cases. In Section \ref{sec:demonstration}, we detail the demonstration scenarios. %for \cynetdiff{}.

\section{Preliminaries}
\label{sec:preliminaries}

In this section, we give formal statements for the network diffusion models and the influence maximization problem.

\paragraph{Diffusion Models.} Diffusion models describe how the cascade takes place in a social network. %For this work, we focus on the \textit{independent cascade} (IC) \cite{goldenberg2001talk, goldenberg2001using}, although \cynetdiff{} can simulate the \textit{linear threshold} (LT) \cite{granovetter1978threshold, schelling2006micromotives} model as well. 
In \textit{linear threshold} (LT) model, given a (possibly directed) graph $G=(V, E)$, the process starts at time $0$ with an initial set of active nodes $S$, called the \textit{seed set}. When a node $v \in S$ first becomes active at time $t$, it will be given a single chance to activate each currently inactive neighbor $w$. The activation succeeds with probability $p_{v,w}$ (independent of the history thus far). If $w$ has multiple newly activated neighbors, their attempts occur in an arbitrary order. If $v$ succeeds, then $w$ will become active at time $t+1$; but whether or not $v$ succeeds, it cannot make any further attempts to activate $w$ in subsequent rounds. The process runs until no further activation is possible.  In the \textit{linear threshold} (LT) model, given a (possibly directed) graph $G=(V, E)$, a node $v$ is influenced by each neighbor $w$ according to a weight $p_{v,w}$ such that $\sum_{w \in \partial v} p_{v,w} \le 1$, where $\partial v$ represents the set of (in-)neighbors of $v$. Each node $v$ chooses a \textit{threshold} $\theta_v$ uniformly from the interval $[0,1]$; this represents the weighted fraction of $v$'s neighbors that must become active for $v$ to become active. The process starts with a random choice of thresholds for the nodes, and an initial set of active nodes $S$, called the \textit{seed set}. In step $t$, all nodes that were active in step $t-1$ remain active, and we activate any node $v$ for which the total weight of its active neighbors is at least $\theta_v$. The process runs until no more activation is possible.

\paragraph{Influence Maximization (IM)}

For a given model, let $\sigma(S)$ denote the expected number of nodes activated after running the diffusion process to completion with initial seed set $S$. Given a diffusion model and a budget $k$, we wish to choose the set $S$ such that $\abs{S} = k$ and $\sigma(S)$ is maximized. Notably, for both the IC and LT models, Kempe et al. \cite{kempe2003maximizing} showed that the influence function $\sigma(\cdot)$ is submodular, and thus its maximum value can be approximated with the greedy algorithm \cite{nemhauser1978analysis}. The greedy algorithm for this problem is computationally expensive, as it requires a large number of evaluations of the influence function $\sigma(\cdot)$. To improve this, the CELF algorithm \cite{leskovec2007cost} was introduced as an optimized version of the greedy algorithm, requiring fewer evaluations of $\sigma(\cdot)$. Despite this optimization, the CELF algorithm still requires a substantial number of evaluations of $\sigma(\cdot)$, and optimizing the speed of these evaluations is a point of focus for implementations of this algorithm.  

%\section{Methodology} Other than everything else, there should be a subsection on run time and space gains. 1 page
% Replacing Methodology with a section on CyNetDiff (since the content is the same).

\section{\cynetdiff{}}
\label{sec:cynetdiff}

In this section, we detail the technologies and implementation techniques used in \cynetdiff{}---the library introduced in this paper.

\paragraph{Cython.}

The desire for greater speed and lower memory usage immediately suggests the use of a compiled language instead of an interpreted one like Python. However, we would like our software package to maintain the greater flexibility provided by  Python and its ecosystem. To accomplish both of these goals, we wrote the performance-critical portions of our library in Cython \cite{bbcdss-cython-11} and included some supporting Python utilities.

Cython is a Python language extension that allows for compilation in C and C++ while still providing code callable from Python. This made Cython the ideal technology for providing a high-level Python interface with similar performance to a compiled language.

\paragraph{Data Structures}

To take full advantage of the additional performance provided by Cython, we represent graphs within the library using array-based data structures tailored to lower-level languages. These data structures have lower memory overhead and allow for faster execution time by Cython.

We opted to store the underlying graphs in the \textit{compressed sparse row} 
(CSR) format \cite{k-csrfrg-20}, using the built-in Cython \texttt{array} data 
structure. At a high level, this format stores the out-neighbors for each node 
in contiguous memory (unlike an adjacency list, which uses pointers), with an 
additional indexing array indicating where the neighbors for each node start.

Although the CSR format makes it difficult to modify the graph once it is stored, it has a lower memory footprint than the adjacency list, and allows efficient queries for the outgoing neighbors of a node without the need for pointer lookups. Thus, CSR format is conducive to efficient, repeated traversals, making it ideal for the internal graph representation used by the library.

We also provide utility functions for the creation of model classes directly from NetworkX graphs. These functions convert NetworkX graphs into the CSR format, instantiate the corresponding model, and return the resulting model class to the client code. This has a substantial impact on the usability of the package, as NetworkX is a well-established library for graph analysis tasks, allowing for easy integration of \cynetdiff{} into existing research pipelines. This makes \cynetdiff{} an effective drop-in replacement for \ndlib{}.

\paragraph{Algorithms}

To facilitate an efficient implementation, we need the following folklore observation about the locality of node activation.

\begin{observation}
    In both the IC and LT models, any node $v$ activated at time $t$ must have at least one in-neighbor $u$ activated at time $t-1$, unless $v$ is a seed node. 
\end{observation}

This immediately suggests that the newly activated nodes in each iteration can be
determined from the out-neighbors of the activated nodes from the previous iteration.
We applied this observation in our implementation of these models, as we use a BFS-based traversal algorithm to determine which nodes are activated in each iteration. As a result, the work performed during the simulation of these models by \cynetdiff{} is proportional to the number of edges incident to
activated nodes. This can be much smaller than the size of the
entire graph when the number of seed nodes is small.

This optimization is very important for improving the runtime in workloads
like the CELF algorithm, where many simulations have very few nodes activated. 
This is especially the case at the beginning of the algorithm's execution, 
since the marginal gains for every individual seed must be computed, and 
only a small portion of the graph is traversed in each iteration as a result.

% TODO does there need to be more?

\section{Demonstration}
\label{sec:demonstration}

%1.5 pages
%Experimental Setup 
%1. Benchmarks - thorough
%2. Visualizations - software aspects
%3. Influence Maximization
% NOTE notebook demo used here: https://kexinrong.github.io/papers/eva-demo-vldb23.pdf

We describe the demonstration scenarios for \cynetdiff{}. As this package
is intended to augment existing research applications in Python, the demo will focus on
how the package integrates into the Python ecosystem for data analysis, and how it handles the increased scale of datasets. These will be conducted within a Jupyter notebook. For brevity, we focus on the IC model only for our demonstration, although \cynetdiff{} can simulate the LT model. 

\begin{figure*}[t]
\centering
\begin{minipage}{.4\textwidth}
  \centering
  \resizebox{\textwidth}{!}{\includegraphics{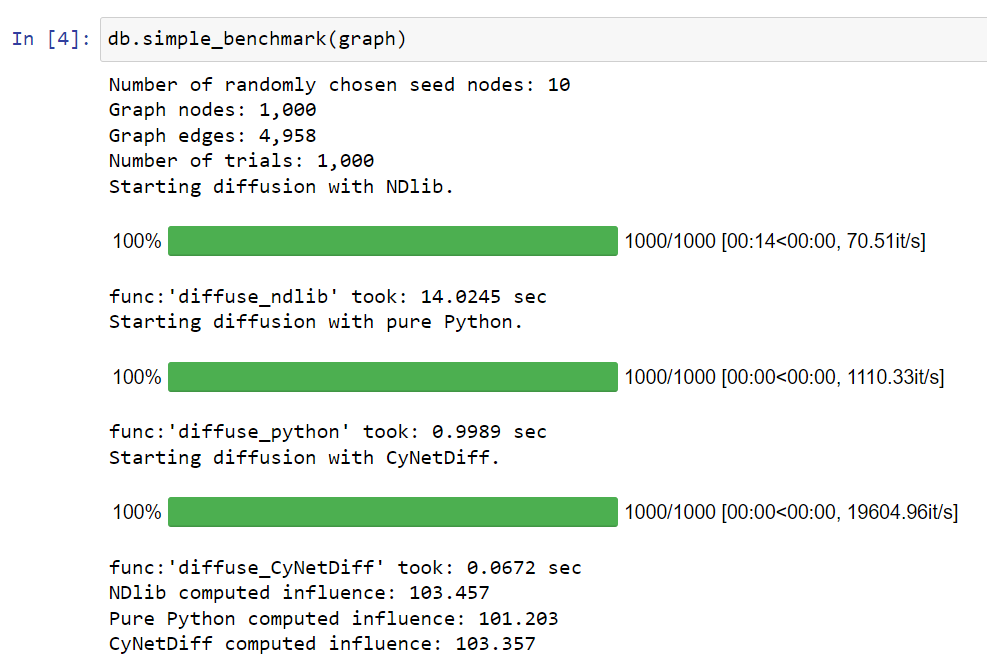}}
  \subcaption{Benchmark output on a synthetic random graph.} 
  \label{fig:benchmark}
\end{minipage}\hspace{.35cm}
%\vspace{.25in}
\begin{minipage}{.55\textwidth}
  \centering
  \resizebox{\textwidth}{!}{\includegraphics{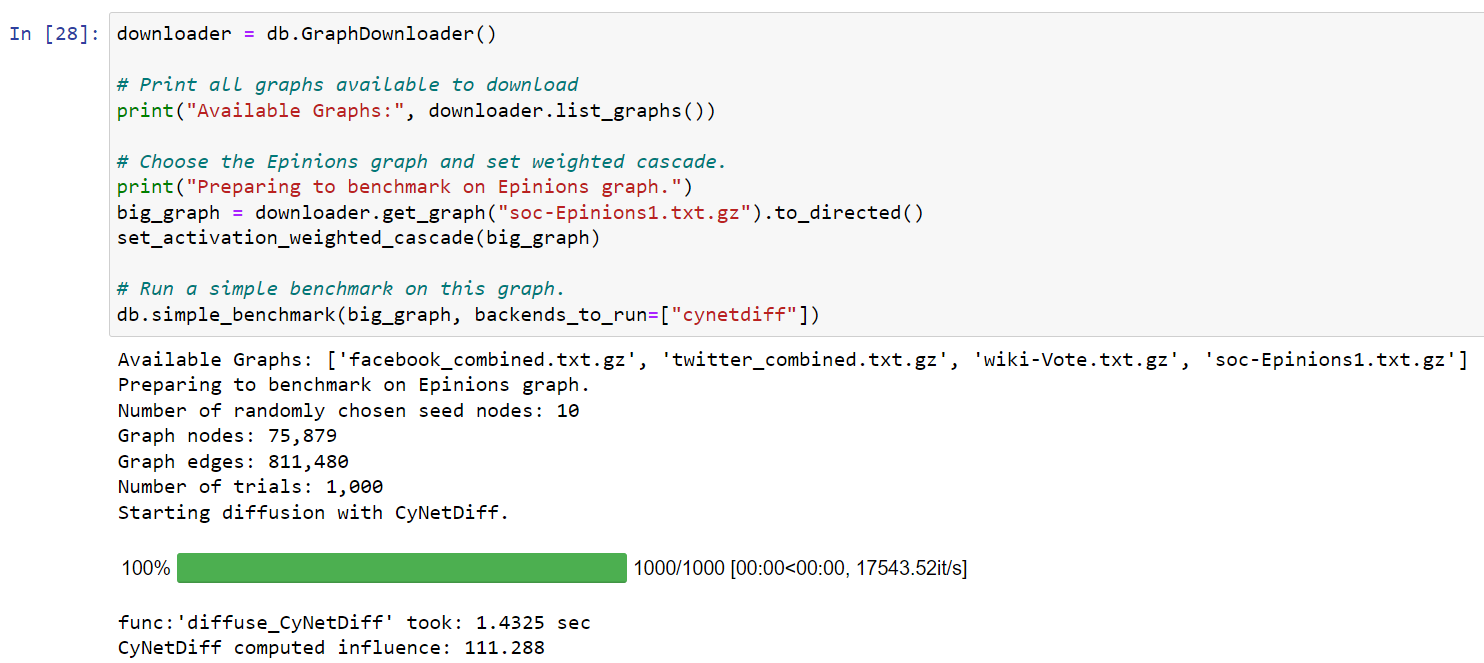}}
  \subcaption{Benchmark output on a real-world graph.}
  \label{fig:benchmark_snap}
\end{minipage}
% \begin{minipage}{.34\textwidth}
%   \centering
%   \resizebox{\textwidth}{!}{\input{figures/diffusion_graph}}
%   \subcaption{Wikipedia network}
% \end{minipage}\hspace{.5cm}
% \begin{minipage}{.34\textwidth}
%   \centering
%   \resizebox{\textwidth}{!}{\input{figures/diffusion_graph}}
%   \subcaption{Epinions network}
% \end{minipage}
\caption{Outputs from \texttt{simple\_benchmark} in the demonstration.
The \texttt{it/s} in the output refers to the number of independent model simulations executed per second.} %\label{fig:ic-wc}
\end{figure*}

% \begin{figure*}[h]
%     \centering
%     \begin{subfigure}{0.45\textwidth}
%         \centering
%         \includegraphics[width=\linewidth]{figures/simple_benchmark_screenshot.jpg}
%         \caption{Output from \texttt{simple\_benchmark} in the demonstration, running on a randomly generated graph.}
%         \label{fig:benchmark}
%     \end{subfigure}
%     \hfill
%     \begin{subfigure}{0.45\textwidth}
%         \centering
%         \includegraphics[width=\linewidth]{figures/simple_benchmark_screenshot_real.jpg}
%         \caption{Running \texttt{simple\_benchmark} running on a real-world graph.}
%         \label{fig:benchmark_snap}
%     \end{subfigure}
% \end{figure*}

\paragraph{Benchmarks}

The first scenario guides participants through interactive benchmarks that introduce the IC model, and compare the performance of \cynetdiff{} with other implementations for these tasks. This is done through the introduction of the \texttt{simple\_benchmark}
function, which can be used to run comparative benchmarks for network diffusion on an
arbitrary input graph. This function outputs performance information for different diffusion implementations over a configurable number of trials. This will allow participants to easily benchmark additional graphs of their choosing and observe the performance of \cynetdiff{}. Output from running this is shown in Figure \ref{fig:benchmark}. The benchmarks in this section cover both synthetic data (generated using graph generation functions provided by NetworkX) and real-world data obtained from SNAP \cite{snapnets}. Participants are encouraged to experiment with different benchmark parameters and observe the effect this has on the performance of the different implementations. An example of a benchmark run on a real-world dataset is shown in Figure \ref{fig:benchmark_snap}.

We next provide some sample benchmarks in Table \ref{tab:run_times}. %Some sample benchmarks conducted with the code that the participants will use are included in Table \ref{tab:run_times}. 
These benchmarks were conducted on a Dell Precision Tower 5810 with a 3.0 GHz Intel Xeon E5-1660 v3 processor. The three implementations were the \cynetdiff{} library, \ndlib{} library, and a fast pure Python implementation of the diffusion model written for comparison purposes. The benchmarks consisted of running the IC model $1,000$ times over the input graph using different edge-weight models (EWM). The first is the \textit{trivalency} (TV) model \cite{goyal2011data}, where each edge-weight was drawn uniformly at random from a small set of constants $\{0.1, 0.01, 0.001\}$. The second is the \textit{uniformly random} (UR) model, where the edge-weight was drawn uniformly at random in the interval $[0,1]$. The third is the \textit{weighted cascade} (WC) model \cite{kempe2003maximizing}, where for each node $v \in V$, the weight of each edge entering $v$ was set to $1/\text{in-degree}(v)$. In all of these weighting schemes, undirected edges in the graph were treated as two directed edges. We used two synthetic networks and a real-world network for our benchmarks.
%The first two graphs were a synthetic workload created using random graph generators provided by Network. 
An Erd\H{o}s-R\'{e}nyi graph with parameters $n = 2,000$ nodes, $p = 0.002$, leading to $4,018$ edges. A Watts-Strogatz Small-World graph with parameters $n = 10,000$ nodes, $k = 10$, $p = 0.007$, leading to $50,000$ edges. More details about these graph models and corresponding parameters can be found in the NetworkX documentation. The Facebook graph was obtained from the SNAP dataset and has $4,039$ nodes and $88,234$ edges.

\paragraph{Visualizations}
The next part of the demonstration focuses on visualizations that can be created using \cynetdiff{}. The graphics are created using other libraries with data provided from simulations run with \cynetdiff{}. Each of these visualizations is created in
real-time with reasonably large graphs and a large number of simulations. This demonstrates how the speed of \cynetdiff{} can be used to enhance different applications, as the creation of these visualizations would not be possible using a slower library.

\begin{figure}[h]
  \centering
  \includegraphics[width=7cm,height=6cm]{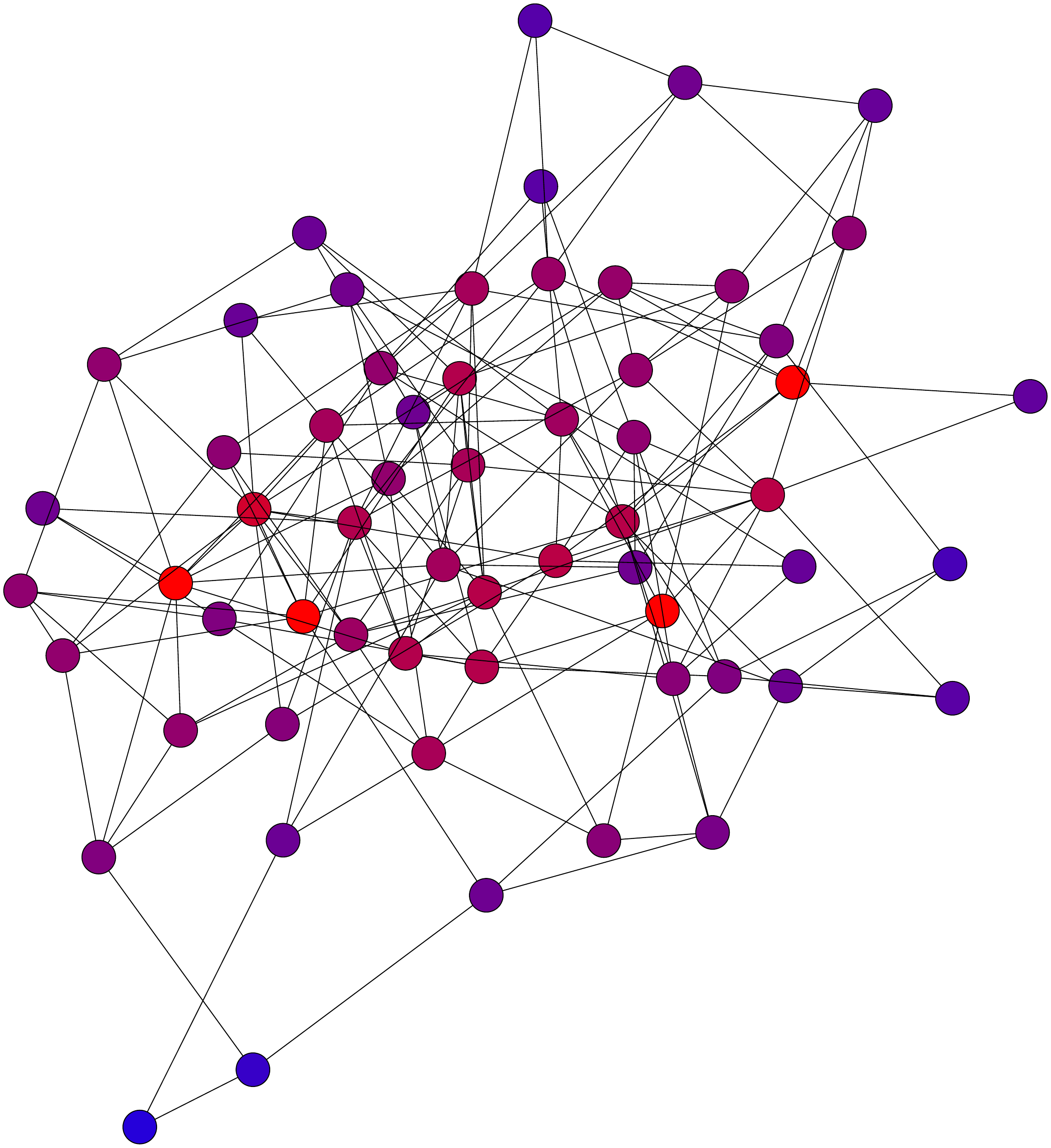}
  \caption{A heatmap created by running \cynetdiff{} and coloring nodes based on how often they were activated. The seed nodes appear completely red, as they were always active.}
  \label{fig:heatmap}
\end{figure}

The first visualization we create is a \textit{heatmap} based on how many times a node was activated across many simulations. The demonstration provides an example of how to run simulations and create a graph with the results, where a node displays redder if it has been activated in more simulations as seen in Figure \ref{fig:heatmap}.

%\label{sec:activation_over_time}
As a second visualization, we use \textsc{Matplotlib} to plot the \textit{mean number of activated nodes over time} for different initial seeds. Each of these plots shows the mean number of activated nodes in each iteration across $1,000$ independent simulations. As before, participants are given multiple scenarios and encouraged to experiment with different parameters to see how this affects the output.

% \begin{figure}
%   \centering
%   \includesvg[width=\linewidth, height=10cm]{figures/diffusion_graph.svg}
%   \caption{Mean number of nodes activated over time over the same randomly generated graph. The plot shows the difference in
%   activation based on the method used to choose a set of seed nodes.}
%   \label{fig:diffusion_graph}
% \end{figure}

\begin{figure}[t]
\centering
%\begin{minipage}{.34\textwidth}
%  \centering
  \resizebox{.35\textwidth}{!}{
  % This file was created by tikzplotlib v0.9.8.
\begin{tikzpicture}

\definecolor{color0}{rgb}{0.12156862745098,0.466666666666667,0.705882352941177}
\definecolor{color1}{rgb}{1,0.498039215686275,0.0549019607843137}
\definecolor{color2}{rgb}{0.172549019607843,0.627450980392157,0.172549019607843}
\definecolor{color3}{rgb}{0.83921568627451,0.152941176470588,0.156862745098039}
\definecolor{color4}{rgb}{0.580392156862745,0.403921568627451,0.741176470588235}
\definecolor{color5}{rgb}{0.549019607843137,0.337254901960784,0.294117647058824}

\begin{axis}[
legend cell align={left},
reverse legend,
legend style={
  fill opacity=0.8,
  draw opacity=1,
  text opacity=1,
  at={(0.98,0.13)},
  anchor=east,
  draw=white!80!black
},
tick align=outside,
tick pos=left,
x grid style={white!69.0196078431373!black},
xlabel={Iteration %\(\displaystyle k\)
},
xmajorgrids,
xmin=-1, xmax=8,
xtick style={color=black},
y grid style={white!69.0196078431373!black},
ylabel={Mean number of nodes activated},
ymajorgrids,
ymin=9, ymax=20,
ytick style={color=black}
]

\addplot [thick, color4, mark=*, mark size=1, mark options={solid}]
table {%
0 10.000
1 16.231
2 17.625
3 17.956
4 18.031
5 18.045
6 18.048
7 18.048
};
\addlegendentry{CELF}
\addplot [thick, color5, mark=*, mark size=1, mark options={solid}]
table {%
0 10.000
1 11.756
2 12.119
3 12.206
4 12.219
5 12.222
6 12.223
7 12.223
};
\addlegendentry{Random Sampling}

\addplot [thick, color3, mark=*, mark size=1, mark options={solid}]
table {%
0 10.000
1 13.019
2 13.677
3 13.824
4 13.841
5 13.846
6 13.847
7 13.847
};
\addlegendentry{Highest Degree}
\end{axis}

\end{tikzpicture}
  }
%  \subcaption{Simulated random graph.}
%  \label{fig:diffusion_graph}
%\end{minipage}\hspace{.5cm}
%\vspace{.15in}
%\begin{minipage}{.34\textwidth}
%  \centering
%  \resizebox{\textwidth}{!}{\input{figures/diffusion_graph}}
%  \subcaption{Real-world graph.}
%\end{minipage}
% \begin{minipage}{.34\textwidth}
%   \centering
%   \resizebox{\textwidth}{!}{\input{figures/diffusion_graph}}
%   \subcaption{Wikipedia network}
% \end{minipage}\hspace{.5cm}
% \begin{minipage}{.34\textwidth}
%   \centering
%   \resizebox{\textwidth}{!}{\input{figures/diffusion_graph}}
%   \subcaption{Epinions network}
% \end{minipage}
\caption{Mean number of nodes activated over time. The data for this plot was dynamically generated during the demonstration on a
random regular graph generated by NetworkX.}
\label{fig:celf_graph}
\end{figure}
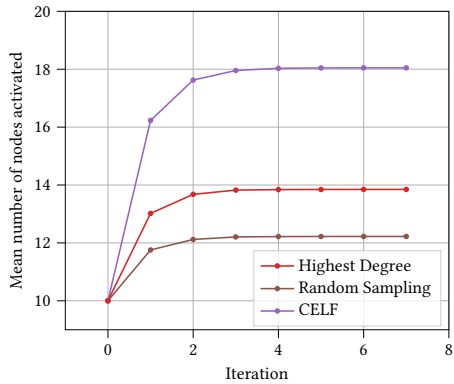

\paragraph{Influence Maximization}
The final scenario in the demonstration focuses on the task of influence maximization. To demonstrate the performance of our implementation of network diffusion, we implement the CELF \cite{leskovec2007cost} algorithm using all of the network diffusion implementations mentioned here as backends. We then compare the influence of the seed set generated by this algorithm to that of the other methods using the plotting functions introduced earlier. %in Section \ref{sec:activation_over_time}.

Participants are encouraged to try different backend implementations for this algorithm, but \cynetdiff{} is used in the demonstration to keep a reasonable runtime. The plot generated by this is shown in Figure \ref{fig:celf_graph}. 

For completeness, we also include a sample comparative benchmark for the CELF
algorithm, see Table \ref{tab:celf_times}. This was performed on a random
7-regular (each node has degree 7) graph with $5,000$ nodes and $35,000$ edges, generated by NetworkX.

\begin{table}[t] 
    \centering
	\begin{tabular}{p{2cm} c c r r}
		\hline \hline
		Graph & EWM & \cynetdiff{} & pure Python & \ndlib{} \\
		\hline
        \multirow{3}{2cm}{Erd\H{o}s-R\'{e}nyi} 
%            & Trivalency       & 1.000 & 11.148 & 193.979\\
%        & & & Uniformly Random & 1.000 & 11.733 & 203.434\\
%        & & & Weighted Cascade & 1.000 & 11.305 & 198.176\\
        & TV       & 1 & 11 & 194\\
        & UR & 1 & 12 & 203\\
        & WC & 1 & 11 & 198\\
        \hline
        \multirow{3}{2cm}{Watts-Strogatz}
%            & Trivalency       & 1.000 &  8.686 & 283.391\\
%        & & & Uniformly Random & 1.000 & 11.007 & 327.225\\
%        & & & Weighted Cascade & 1.000 &  8.819 & 311.618\\
        & TV       & 1 &  9 & 283\\
        & UR & 1 & 11 & 327\\
        & WC & 1 &  9 & 312\\
        \hline
        \multirow{3}{1em}{Facebook}
%            & Trivalency       & 1.000 &  8.299 & 81.336\\
%        & & & Uniformly Random & 1.000 & 12.119 & 44.688\\
%        & & & Weighted Cascade & 1.000 &  7.864 & 71.335\\
        & TV       & 1 &  8 & 81\\
        & UR & 1 & 12 & 45\\
        & WC & 1 &  8 & 71\\
        \hline\\
	\end{tabular}
 	\caption{Comparison of run-times for independent cascade run with 100 seeds on different graphs. Runtimes are normalized and rounded over
    each row so that the fastest benchmark in each row is $1$.}
    \label{tab:run_times}
    \vspace{-.5cm}
\end{table}

\begin{table}[t] 
    \centering
	\begin{tabular}{l c r r}
		\hline \hline
		Graph & EWM & \cynetdiff{} & pure Python \\
		\hline
        \multirow{2}{10em}{Random $7$-regular}
        & TV & 2 & 26\\
        & WC & 10 & 153\\
        \hline\\
	\end{tabular}
 	\caption{Comparison of run-times for the CELF algorithm run with 10 seeds. Runtimes are in seconds. Results for \ndlib{} are not reported because they did not finish within 5 minutes.}
    \label{tab:celf_times}
    \vspace{-.75cm}
\end{table}

\section{Conclusion and Future Work} % + Ref. 1/2 page

In this work, we have described and demonstrated the effectiveness of \cynetdiff{} on a variety network diffusion tasks. The demonstration scenarios and benchmarks show how the speed of the library enables running larger experiments and the creation of interesting visualizations. This will enable faster and more comprehensive
experiments going forward by researchers studying network diffusion
and influence maximization.

There are several avenues for further work. In one direction, there is potential to continue to improve the performance of \cynetdiff{} by adding parallelism. In another, introducing this library enables performing larger-scale experiments, enabling an expansion of the
scope of experiments conducted in prior work.

%\clearpage

\bibliographystyle{ACM-Reference-Format}
\bibliography{refs.bib}
\end{document}